\documentclass[manuscript]{aastex}

\slugcomment{Submitted to Astrophysical Journal (2010)}

\shorttitle{Bright points in and around sunspots}
\shortauthors{Choudhary et al.}

\begin{document}


\title{Sunspot Bright Points}


\author{Debi Prasad Choudhary\altaffilmark{1,2} and Toshifumi Shimizu\altaffilmark{1}}
\affil{Institute of Space and Astronautical Science, Aerospace Exploration Agency (ISAS/JAXA)
   3-1-1 Yoshinodai, Sagamihara, Kanagawa 229-8510, Japan}

\altaffiltext{1}{Institute of Space and Astronautical Science, Japan Aerospace Exploration Agency (ISAS/JAXA)
   3-1-1 Yoshinodai, Sagamihara, Kanagawa 229-8510, Japan
       E-mail shimizu@solar.isas.jaxa.jp}
\altaffiltext{2}{Department of Physics and Astronomy, California State University Northridge, 18111 Nordhoff Street, 
       Northridge, CA, 91330}

\begin{abstract}
We used the flux calibrated images through the Broad Band Filter Imager and Stokes Polarimeter data obtained with the Solar Optical Telescope onboard the Hinode spacecraft to study the properties of bright points in and around the sunspots.  The well isolated bright points were selected and classified as umbral dot, peripheral umbral dot, penumbral grains and G-band bright point depending on their location. Most of the bright points are smaller than about 150 km. The larger points are mostly associated with the penumbral features. The bright points are not uniformly distributed over the umbra but preferentially located around the penumbral boundary and in the fast decaying parts of umbra. The color temperature of the bright points, derived using the continuum irradiance, are in the range of 4600 K to 6600 K with cooler ones located in the umbra. The temperature increases as a function of distance from the center to outside. The G-band, CN-band and CaII H flux of the bright points as a function of their blue band brightness increase continuously in a nonlinear fashion unlike their red and green counterpart. This is consistent with a model in which the localized heating of the flux tube deplete the molecular concentration resulting the reduced opacity which leads to the exposition of deeper and hotter layers. The scatter in CaII H irradiance is higher compared to the G-band and CN-band irradiance. The light curve of the bright points show that the enhanced brightness at these locations last for about 15 to 60 minutes with the least contrast for the points out side the spot. The umbral dots near the penumbral boundary are associated with elongated filamentary structures.

The spectropolarimeter observations show that the G-band brightness closely follows their magnetic filling factor. As the filling factor decreases the brightness increases. Generally, the umbral dots have higher magnetic field and larger Doppler velocity compared to their counterpart outside the spot. The ``flux tubes" in the umbra are less inclined compared to the ones associated with the penumbra.

We discuss these observation in the framework of sunspot models that include the creation of umbral dots by the plasma trapped in between the magnetic flux tubes. The upward intruding plasma through the nonmagnetic columns between the fluxtubes of sunspot produce the bright points and heat the matter inside of adjacent tubes. The heated plasma flows in the direction of reduced gas pressure. Similar localized heating of penumbra leads to the origin of penumbral grains. The G-band bright points are mostly the vertical flux tubes embedded in photospheric material. The varying degree of G-dand, CN-band and CaII H brightness of these points are due to the combined effect of different degree of heating and exposure of hot flux tube walls for viewing deeper photospheric layers.

\end{abstract}



\section{Introduction}

The compact and time varying bright regions located in the form of umbral dots, light bridges and penumbral features harbor signatures of underlying physical processes below the visible layers \citep{par79b,spr06,sch06,cho86}. Last two decades of ground based high spatial resolution observations have revealed that umbra is inhomogeneous divided by light bridges and filled with bright points and elongated structures that can not be primarily produced by granular convection processes \citep{sob93,liv91,rim08}. The penumbra is more dynamic with stratified temperature, velocity and magnetic field structure and consists of elongated channels with intermittent Evershed flow \citep{rou02,wes01a,wes01b,rou03,rim94}. Extremely bright grains and regions of hot upflows are also observed at the middle of penumbra \citep{den08,bec08}. The features are integral part of the magnetic structure, which could be aggregation of many separate flux tubes pressed together to form a flux bundle or monolithic large flux tube \citep{del01}. 

Theoretically, there are three mechanisms that can produce bright points with some of the observed properties in the sunspot. One of them is primarily based on the properties of fluids with high electrical conductivity trapped in magnetic field described by \citet{cha61} that were also observed in laboratory systems \citep{nak57}. In case of sunspots, the plasma is trapped between the flux tubes in a cluster. \citet{par79a} proposed that the sunspots are dynamical clustering of many separate flux tubes pressed together as single flux bundle at the visible surface keeping their separate identity beneath. While the magnetic pressure at the center of the flux tube axis is smaller by about 0.5 to 0.75 compared to the boundary near the photospheric layers, the field is more concentrated below. The space between the flux tubes below the photosphere are filled with the field free plasma which become unstable and subjected to oscillation appearing as umbral dots \citep{par79b}. Such oscillatory plasma can also penetrate the surface with velocities of about 270 m s$^{-1}$ \citep{cho86}. The observations of formation and decay process of the sunspots appears to be consistent with these theories of penetrative convection \citep{gar87}. The second category of physical processes are developed for monolithic sunspots as a result of non-linear magnetic convection leading to oscillatory plasma columns \citep{kno84}. The simulations show that the  monolithic sunspots models can also explain, the formations of umbral dots with magnetic fields of 3.8 kilo Gauss, central upward flowing plasma of upto 4 km s$^{-1}$, surrounded by downward motion \citep{sch06}. The velocity field at the top of plasma columns in these processes would display vertical velocities up to 1 to 2 km s$^{-1}$. The third type of theoretical model consider magnetohydrodynamic consequences of thin vertical gas column embaded in a sunspot umbra \citep{deg93a,deg93b}. Geometrically, this is a converging gas column with a week field than that of the ambient medium. In this case, both up and down flowing plasma exists depending on the temperature and density at the top of the column. While the appearance of near circular structure of the compact bright features are consistent with the theoretical models, other properties such as mass motion and magnetic field configuration needs to be examined for a better understanding. Several observations show not all the umbral dots are circular and the associated velocities vary. The penumbra consists of elongated horizontal flux tubes through which the intermittent Evershed flow occurs. Some of the theoretical models describe the penumbra as a result of the returning flux tubes anchored by the inter-granular lane through turbulent pumping \citep{tho02}. Other models interpret penumbral fine structures as due to the convection of field-free radially aligned gaps just below the visible surface \citep{spr06,sch06}.

These different competing models can now be examined in detail with the availability of high spatial resolution observations. With the high resolution spectropolarimetry observations and modeling it has been possible to look for finer descripencies between the theory and obserrvations. For example the TIP and POLIS observations show that the penumbral models with flux tubes embedded in non-magnetic plasma needs to account for the flow \citep{bec08}. The high resolution images of sunspots obtained with adoptive optics system at Dunn Solar Telescope show elongated structures associated with umbral dots towards the penumbral boundary that can not be entirely due to magnetoconvections \citep{rim08}. The imaging and spectropolrimetry observations with the Solar Optical Telescope (SOT) on Hinode provides best possible high spatial resolution observations to study these structures without seeing interruptions. These observations have already contributed to our understanding of the sunspot bright points \citep{wat09a,rie08a}. The similarity of penumbral grains and peripheral umbral dots have been found by Sobotka and Puschmann (2009). On the other hand, there is a difference in the evolution of these two structures \citep{rie08a}. 

In this paper, we use the broad band images and spectropolarimetry observations of sunspots to study the thermal and magnetic properties of isolated bright points in and around sunspots. We find that there is continuity in the properties of bright features from the umbral dots that exists deep inside the umbra to the bright points in the vicinity of sunspots. We discuss these results with contemporary models through a simple cartoon.
 
\section{Observations and Analysis}
We have used the images of the sunspots with Focal Plane Package of Solar Optical Telescope (SOT) on the Hinode Spacecraft \citep{tsu08, kos07}. The sunspot images through Broadband Filter Imager (BFI) with spatial resolution resulting from 1 $\times$ 1 and 4 $\times$ 4 summing were used for this study. After initial processing of the images for removing instrumental affects with standard procedures provided by Hinode team and co-alignment, we perform flux calibration. In order to convert the data number into the units of solar irradiance, we compared the data numbers of solar flux obtained by the Charge Coupled Device on ground and space before and after the launch of Hinode to check the consistency of detector performance. We measured average data numbers of area covering 100 square pixels in relatively non-plage locations of quiet sun away from the sunspots that may represent the average solar flux measured on ground \citep{shi07}. The ground based measurements were performed by observing the disk center and averaging the entire field of view. While the ground measurements may contain some magnetically active locations, we have excluded them from space measurements. Figure 1 presents the ground and space data number comparison for 29 April 2007. We find the general trend of data numbers of the detector remains same both on ground and on space with somewhat larger scatter through the filters with shorter wavelength. Possible inclusion of pixels with magnetic elements in the ground based data could result differences especially in CaII and G-band flux. After correcting for the reduced filter transmission for the observing day, we compare the quiet sun values with a standard zero airmass solar spectrum to obtain the calibration factor to convert the data numbers into solar flux units \citep{weh85}. The flux calibrated images were used to obtain the intensity of bright points in the sunspots. 

The sunspots that were used for this study are given in Table 1. The bright features representing umbral dots, peripheral umbral dots, penumbral grains and bright points in the vicinity of sunspots were identified from the ratio images obtained by dividing G-band and CaII H frames as shown in Figure 2. The contrast of the bright points in the inner portion of sunspots becomes much higher leading to their easy detectability. The higher contrast in the ratio images are due to the featureless umbra in CaII H images which represent the solar atmospheric height of about 150 Km above the G-band height formation \citep{car07}. Most of the sunspot flux tubes, that result the moving magnetic features, do not reach at lower atmospheric heights below about 300 km leading to their non-appearance at lower chromosphere \citep{cho07}. In this case, the CaII H images act as high pass filter enhancing the features that do not reach the heights of about 150 km. Identification of bright features by comparing G-band and CaII H images preclude the points that might be effected due to the presence of brighter locations in close proximity.

The photometry of the bright points were performed following method similar to the crowded-field photometry known as DAOPHOT (Stetson, 1987). The main steps are identification and flux measurement by fitting a Gaussian surface function to the bright point. Initially, the coordinate of the bright points were visually determined from the ratio images. The coordinates match very well with the locations of well defined, isolated bright points found by adopting the automatic procedures that compares the adjacent pixels to look for enhancement of local intensities (Riethmuller et al, 1008, Watanabe et al, 2009). The additional constraint for selecting the bright point for our analysis is successful fitting of Gaussian surface leading to a convergence for photometry flux extraction. The exact center of the bright points were determined by using a derivative search at which the local derivative in the neighborhood of the manually found positions of brightness goes to zero. A 20 $\times$ 20 pixel image around the center was extracted for flux measurements. The image was fitted with a two dimensional elliptical Gaussian function given in Equation (1), around the centroid of the rectilinearly gridded bright point image. 

$ F = A_0 + A_1 e^{-\frac{U}{2}}$ ~~~~~~~~~~~~~~~~~~~~~~~~~~~~~~~~~~~~~~~~~~~~~~~~~~~~(1)

where $U =(\frac{X}{a})^2 + (\frac{Y}{b})^2$. The X, Y are the center and a, b are the axis of the ellipse along the line joining to the centroid of the large portion of the sunspot umbra. The flux was calculated by summing the intensity of the pixels within half width of the Gaussian Surface. Average size of the bright point was computed by taking average of both axis. Figure 3 show typical bright points and corresponding Gaussian surface. Figure 4 shows the histogram of the bright point radius. The extracted flux over the Gaussian function half-width from the co-aligned images through the filters represents the brightness of the different features in the sunspot that are shown in Figures 5. Figure 6 show the continuum flux superposed with the Planck function for different temperatures.  

The G-band images over a period of 5 hours were co-aligned  using the adjacent image shifts to determine their light curve. The integrated flux was measured following the same procedure. Since the extracted image for Gaussian surface fitting is larger with 20 pixels, the inward migration of umbral dots, which is at most about three pixels, does not contribute to the flux measurement. Figure 7 show the typical light curves of various locations.

The spectropolarimeter (SP) observations of the sunspot close to the imaging observations, obtained on 28 April 2007 beginning at 09:00 UT, are used to determine the magnetic and velocity properties of the bright features. By using the inversion of radiative transfer equation for all components of polarized spectrum, we obtained the magnetic and velocity over the sunspot from the SP data.  The inversion code uses Milne-Eddington atmosphere for producing the spectral lines. The magnetic and velocity field maps and the degree of polarization in the spectral lines in the sunspot are shown in Figure 8. The reference velocity (zero velocity) was obtained by averaging the values of the pixels in the umbra that did not contain bright points. We notice the enhanced magnetic and velocity compared to their surrounding. We obtained the magnetic field and velocity of several common identified points of images by following a similar feature of flux extraction by two dimensional Gaussian function fitting. The compact bright points in the continuum image of 6301.5 \AA\/ were close to the coordinates of bright points in the red continuum image. Figure 9 show the filling factor of the bright points that are common in SP and imaging observations plotted along with the G-band bright point intensities in the same scale. Other properties such as magnetic field, Doppler velocity, field inclination and degree of polarization of the bright points derived from the SP observations are shown in Figure 10.

\section{Results and Discussion}

There are two classes of theoretical sunspot structures that can be used to build contemporary models to explain the formation of Umbral Dots. One of them is developed with the hypothesis that the subsurface magnetic field of a sunspot contains several separate flux tubes filled with magnetic field free fluid between the gaps. Through penetrative convection the field free gas oscillates or occasionally penetrates through the surface appearing as bright points \citep{par79b,cho86,gar87}. The other models umbral dots as the result of rising plumes triggered by the oscillatory magnetoconvections \citep{sch06}. In both these models the up flowing plasma at the bright dots has higher density with reduced blackbody temperature compared to the quiet sun. In case of penetrative convection the fluid is magnetic field free, where as in the case of magnetoconvection of monolithic sunspots, the umbral dot field is expected to be mostly in the vertical direction. The plasma columns with lower field compared to the surrounding embedded in a flux concentration will also produce the bright points \citep{deg93a}. In the following, we shall discuss our results in context of these theoretical models.

\subsection{Ratio of G-band and CaII H images}

The ratio of sunspot image obtained through G-band filter and CaII H filter is shown in Figure 2 that shows bright points in the umbra, penumbral boundary and in the penumbra. \citet{car07} have shown that for the SOT/BFI G-band filter that includes a broad upper photospheric contribution samples about 74 km, where as CaII H filter samples higher region at the height of 247 km of solar atmosphere. These heights are derived using the quiet sun atmospheric models \citep{fon90}. In sunspots, the atmospheric model will change significantly, the contribution in CaII H would be at a higher atmospheric level compared to that of G-Band. The sunspot atmospheric models show that the density and pressure has a higher gradient in sunspots compared to the quiet sun \citep{sta81,obr88}. The contribution height between the two lines in the umbra might be smaller than 173 km, which is less than the height of the magnetic features obtained earlier \citep{cho07}. In the penumbra, however, the CaII formation height could be higher than in the umbra \citep{din82}. The ratio images show peripheral umbral dots are mostly located at the foot points of the origin of penumbral fibrils and much brighter than the umbral dots. These bright points are also seen to make inward migration with time. However, with in the time gap of images used for the ratio images, which is about 3 seconds, the motion would result the displacement within a pixel of the detector without affecting the contrast of these points. These points may be considered as the penumbral foot points as they are located mostly at the origin of penumbra. We notice most umbral dots are not completely circular but has a slightly asymmetry directed towards the penumbra as shown in Figure 3a. The ratio of elliptical axis of Gaussian surface is about 1.1. Some of the umbral dots, specially located towards the umbral boundary, display elongated filamentary structures pointed towards the penumbra as shown at the location with arrow mark in contrast enhanced image in Figure 2. Such elongated features are reported earlier \citep{rim08}. Similar to the radial filamentary structures discovered by \citet{liv91}, many umbral dots appear to be linked with magnetic structures, contrary to the expectation of theoretical models that consider them only due to the rising plasma columns in sunspots. The dots are not uniformly distributed in the umbra and are more concentrated at that part of the umbra that decayed sooner. We have examined at least 16 sunspot ratio images. In all these sunspots the umbral dots located at the foot points of the penumbra towards the disk center appear brighter compared to their limb side counter part. Such effect would be expected if the bright points represent inclined magnetic structures in which the limb side has lower depth compared to the ones towards the center of the sun. Earlier observations have also showed the similarity between the umbral dots and  peripheral umbral dots and non-uniform distribution within the umbra \citep{sob09a}. These imaging observations suggest that the bright points in the umbra and its periphery may not be only due to the protrusion of non magnetic material from below the visible surface, but are actually parts of rapidly diverging inclined flux tubes similar to the ones that make penumbra.

The ratio images show that the penumbra is divided into two parts at midway between the umbra and quiet sun. The inner penumbra has higher ratio compared to the outer penumbra. This is roughly the location where the inward and outward migration of penumbral grains are observed \citep{den07}. This division of brightness of penumbra show the temperature stratification with height is different in two parts of the sunspot. The temperature distribution leads to the photospheric layer of inner sunspot hotter, whereas in the outer part the lower chromosphere is hotter contributing more flux through the CaII H filters. The penumbral grains are brighter and elongated with the ratio of axis of 1.4 and few with higher values as shown in Figure 3b. Most theoretical models describe penumbra as horizontal flux tubes embedded in field free plasma with gaps between them similar to the Parker's model of subsurface magnetic structure of sunspots \citep{spr06,hei07,rem09}. In these models the bright features are considered as the hotter field free material surrounding the flux tubes carrying the cooler plasma. Some times they are also considered as the view of photosphere through the gaps of penumbral flux tubes as first reported by \citet{moo81a,moo81b}. While, it is not possible to distinguish between the photospheric view and in-situ hot plasma around the flux tube, we notice that the bright features are elongated and aligned at the middle portion of the penumbral features indicating that they are related to the geometry of the flux tubes. There are other mechanisms such as localized reconnections that can heat the magnetized plasma due to the magnetic reconnection processes as primarily observed in CaII H images \citep{kat07}. Such contribution we would observe as dark features in outer penumbra in the ratio images. However, if the reconnection heats the plasma at lower heights in the flux tubes the features would be elongated. The bright points observed in the penumbra are long lived unlike the reconnection events that has shorter life time. The size distribution derived from the ratio images of the bright points is bimodal peaking below 0.2 arc second and near 0.5 arc sec as shown in Figure 4. The smaller bright points are umbral dots similar to the features observed by \citet{rim08}. The larger ones are peripheral umbral dots or penumbral foot points. The sample with large size than 0.5 arc second are the penumbral grains.

\subsection{Bright Point Photometry}

The photometry measurement of the bright point flux through blue (4505 \AA\/), green (5550 \AA\/) and red (6684 \AA\/) continuum filters along with the planck functions for the temperature range of 4600 to 6600 K are shown in Figure 6. The Figure shows that there is a gradual increase of temperature from cool umbral dots to the hotter G-band bright point in the vicinity of the sunspots with in range of 2000 K. Earlier spectroscopic observations showed that the sunspot umbra mainly consists of a two component atmosphere \citep{mak63}. The cold component has temperature of $\sim$ 4000 K, where as the hot component temperature is $\sim$ 5700 K. The hot component occupies about 10\% of the area and contribute to the 50\% of the radiation. Our observations show that while these temperatures are consistent with the hot component umbra derived using the spectroscopic diagnostic, we find that the hotter component actually is due to the umbral dots located at the boundary of penumbral fibrils \citep{mak63}. While the umbral dots are cooler, the penumbral grains and outside bright points are hotter. Consistent with the blackbody character of the bright points, the continuum flux through green and red filter change linearly with respect to the blue continuum flux as shown in Figure 5. We also notice that the G-band brightness of the points surrounding the sunspots represented by the blue plus sign are brighter when located near the solar limb compared to the disk, a trend similar to the quiet sun G-band bright points.

As shown in Figure 5, the flux measured through the filter in molecular bands (CN and G-band) does not show the trend for black body character of the bright points in which case the corresponding values would have been similar to continuum flux trend. Instead of a linear relationship between the blue ($F_b$) and G-band ($F_g$) flux, we find the non linear dependence represented by the dashed line following the Equation 2.

$F_g = (A \times F_b) + F^2_b + B$ ~~~~~~~~~~~~~~~~~~~~~~~~~~~~~~~~~~~~~~~~~~~~~~~~~~~~(2)  

where A and B are constants. The theoretical modeling of G-band bright points are focused to explain the measurements carried out the  quiet sun \citep{ber98}. In quiet sun, the G-band bright points are mostly associated with the magnetic concentrations that has elevated temperature leading to the dissociation of primarily CH molecules. At these wavebands the opacity remains unchanged compared to the neighboring continuum opacity that is due to H$^{-1}$. As a result the deep photospheric layers of hotter flux tubes are exposed through the CH-wavelength window \citep{san01,ste01a,ste01b}. The flux tube geometry plays an important role, where the evacuated magnetic elements for maintaining the pressure balance with the surrounding exposes the hotter layers of photosphere that is enhanced for inclined flux tubes leading to higher brightness near the solar limb compared to their disk counterpart \citep{rut01}. Here also we notice the enhanced G-band brightness of circum-sunspot bright points compared to their disk counterpart (Figure 5a). However, the bright points in the sunspot do not show such disk-to-limb variation trend of the brightness. This behavior may be expected from already inclined flux tubes that would show the G-band brightness with an upper limit. In that case, our observation imply that unlike their quiet sun counterpart, the sunspot bright points would represent inclined flux tubes, which is consistent with the magnetic field inclination values obtained from SP observations shown in Figure 10,. The G-band spectral synthesis with model solar atmosphere show that the G-band intensity does not vary in a linear fashion with respect to the continuum intensity and depends on the magnetic properties of the location \citep{san01}. We observe a similar trend for CN band as shown in Figure 5b. In CaII H filter, we also observe a similar trend with more scatter shown in Figure 5c. In case of CaII H the enhanced ionization in hot environment could result a similar situation as molecular dissociation alter the opacity of the region. The non linear dependence of G-band flux as well as flux measured through CN and CaII H filters could be due to the magnetic properties of the bright locations in the sunspots as discussed later.

\subsection{The Light Curves}

The bright point light curves in Figure 7 show that the large contrast enhancements observed in umbral dots, peripheral umbral dots and umbral grains compared to the circum-sunspot bright points. The life time of bright points are in the range of 15 minutes to about one hour. This is in agreement with several earlier measurements \citep{kus86}. The higher contrast in the umbral dots compared to the G-band bright points could be due to the intense heating of the cold gas in side the spot compared to the ones outside. Inside the sunspot, the material is generally cold, isolated by the magnetic field from the hot plasma in the solar convection zone. The physical mechanism that can heat the material does so for a limited time exposing the hot material underneath. The G-band bright points outside the sunspot are generally embedded in a hot environment. The external heat input does not have a dramatic effect as in case of sunspot material.

\subsection{Magnetic Properties}

The total magnetic flux, Doppler velocity and the degree of polarization map derived from the SP observations are shown in Figure 8. While the magnetic field map show that the bright points corresponding to higher magnetic field compared to the surroundings, the polarization map show enhanced polarization in agreement with earlier results \citep{bec07}. Both these observations indicate the magnetic nature of the bright points. While the magnetic field are enhanced at these location of bright points that appear as blobs in the magnetograms, the field at the center of these blobs appear to be some what lower than the surroundings. The line of sight velocity map show continuous channels of enhanced velocity elongated towards the penumbra starting at some of the bright points. 

The quantitative magnetic properties of the bright points are presented in Figure 9 and 10. The relationship between the magnetic filling factor and the G-band flux of bright points derived from the nearly co-temporal observations of BFI and SP observations are shown in Figure 9 in the same scale. We measure the flux at the bright point in G-band and their corresponding red continuum using the BFI images. Similarly, the flux of the bright points at the continuum wavelength of 6301.5 \AA\/ and their filling factor from inverted SP data are extracted. These are the two different data sets that are used to determine the continuum flux of sunspot bright points in red wavelength, G-band intensity and the magnetic filling factors. Although the spatial coordinates of the bright points in BFI and SP data are close, we are not sure if they are the same objects due to the temporal gap between two observations. Therefore, we plot the G-band flux and magnetic filling factor as a function of red continuum and 6301.5 \AA\/ wavelength respectively to examine the general trend. We notice a remarkably tight relationship between the magnetic filling factor with G-band bright point intensity where the magnetic features with smaller filling factor are brighter. Although the SOT on Hinode has very small scattered light contribution, it would effect most for the umbral dots resulting their lower value indicated by the arrow in Figure 9. The reduced filling factors are expected for the flux tube that "flare out" more \citep{par79b} leading to the exposure of larger cross-section of photosphere devoid of CH molecules.

Figure 10a show that the brighter points are of lower magnetic flux compared to the darker counterpart, which are mostly umbral dots similar to the earlier results \citep{bec69, kne73,soc04,rim04,wat09a,wat09b,wie93}. However, our the Doppler velocity measurements also show smaller values for the brighter spots compared to their darker counterpart in the sunspot (Figure 10b). The brighter points situated in outer penumbra and outside the sunspot also display large downward velocity values as represented by six points shown in a circle in figure 10b. These locations could harbor the Evershed flow material draining along the flux tube.  Although, we shall discuss the velocity properties of the bright sunspot features in a later paper, here we confine the comparison with the velocity measurements having high spatial resolution observations. Several velocity measurements in different wavelength show diverse properties of umbral dots and penumbral grains. While some penumbral grains showed upflows, the umbral dots displayed large down flows \citep{har03}. The relative red shift velocities are observed at these points indicates the presence of down flowing plasma \citep{wie93}. The proper motion studies show the bright features moving towards the umbra \citep{kit86,mal92,rim94}. The ground based high resolution observations using adoptive optics by \citet{rim04} showed the the bright points displaying $\pm$ 300 m s$^{-1}$, with brighter ones near zero velocity. However, the Hinode observations also showed relative blue shifts indicating a upward flow plasma with respect to the surrounding at these locations \citep{wat09a}. The observations with La Palma Stokes Polarimeter show very small upflows with velocity of about 100 m s$^{-1}$.  Hinode SP data inversion also find downflow velocities at higher photosphere \citep{sob09b, rie08a}. The velocity measurements, being relative to the surrounding, they are effected by the choice of reference velocity region. Some of the discrepancy of the velocity measurements could arise due to the selection of the reference region. Our velocity measurements are derived from the Stokes profile inversion that gives Doppler velocity measured from the line shift. The velocity of the umbral region devoid of bright features was taken as the reference, with respect to which all except two penumbral bright points show down-flow velocities. Figure 10c show a wide range of the field inclination for the brighter points with smaller range for umbral dots. The horizontal field are found at these locations in earlier measurements \citep{wie93,lit91}. Combined with the discovery of association of elongated features, the flatter field of these locations indicate the presence of inclined flux tubes at the sunspot bright points. The degree of polarization is higher at these location of bright points with decreasing trend for brighter points as shown in Figure 10d. As the polarized signal is essentially due to the magnetic field, the bright features are likely the site of magnetic flux tubes instead of field free plasma.

\section{The Model}

The observational properties of the sunspot bright points, presented in section 3, show their magnetic structure. While several properties of the bright points agree well with earlier results, our G-band flux measurements of four sunspots, indicate that these locations are showing properties of ``flaring flux tubes" as described by \citet{rut01}. We also notice that careful examination of the umbral bright points in the sunspots show association of elongated filamentary structure extending towards the penumbra similar to the features reported earlier \citep{liv91, rim08}. Following the theory for the umbral dots, originally proposed by \citet{par79b}, we interpret the results with a model as illustrated Figure 11. 

The dark vertical lines with divergence at the top in the umbral dot in Figure 11 a represent the ``Flaring Magnetic Field" above the visible solar surface shown by the dashed line. The diverging magnetic flux tubes have gas pressure decreasing in downward direction as the magnetic field increases leading to higher magnetic pressure. The inter flux tube space is filled with field free hotter plasma, compared to that inside tube, penetrating through the gaps in the subsurface clustered flux tube bundle from the quiet sun. As the plasma column is trapped between the magnetic field, they would be subjected to magnetic instabilities in a similar fashion as described by \citet{par79b}. The hotter oscillating plasma will heat the adjacent flux tubes through heat exchange \citep{ham04,spr81,sch91,jha92}. The heated plasma would eventually flow downwards following the pressure gradient in the flux tubes. However, if these oscillating plasma column were manifesting as the umbral dots, than we would have observed the G-band intensity consistent with the blackbody curve in which the brightness increase depends linearly with blue band intensity. But, since we observe that the G-band intensity is nonlinearly dependent on blue band intensity in a manner that would be expected from the flux tubes as described by \citet{rut01}, we suggest that the G-band bright points correspond to the diverging flux tube locations adjacent to the oscillating magnetic field free plasma column. Since these are the same points that appear brighter in the continuum wave bands, the umbral dots represent the photospheric interaction of the diverging flux tubes that show the magnetic properties described in the previous section.

The penumbral foot points are more inclined and located at the edge that is in contact with the field free convective zone material. The flux tube plasma at these locations will receive more heat flux from the adjacent field free convective zone plasma leading to the heating of a large cross-section of hot wall in the line of sight. This effect would result their higher brightness appearance compared to the bright points inside the umbra. The sunspot has a gradually lower elevation from out side to the center resulting the Wilson depression. The same effect would enable viewing a larger cross-section of hot wall of the flux tube leading to the brighter appearance of sun center side penumbral foot points. In this model the penumbral grains are due to the viewing of a larger cross section of inclined flux tubes embedded in field free plasma \citep{sch06}. The localized heating could arise either by in-situ heating by magnetic reconnection \citep{kat07} or other plasma processes \citep{sch03,rui08}. The G-band bright points in the vicinity of the sunspot are similar to that described in \citep{ber98}.

\section{Summary and Conclusions}

The analysis of high spatial resolution imaging and spectropolarimetry observations of sunspots show the following properties of their bright points. (a) The bright points are non-circular and associated with elongated filamentary structures. (b) The G-band intensity deviates from the expected blackbody temperature.(c) The locations of bright points display magnetic properties. These properties can be best understood if the subsurface structure of sunspots are porous with intersprsed convective zone plasma. 

Understanding the stability and decay of the sunspots is a major problem of modern solar physics. The properties of bright points in and around the sunspot is a crucial part of the sunspot structure playing important role in this processes. Our study shows that the bright points are mostly magnetic concentration compared to their surroundings. The properties are compatible with the description in which the bright points represent the locations inside the diverging flux tubes instead of the inter-flux tube gaps as proposed by \citet{par79b}. If we consider them as purely due to magneto-convection the nonlinear relationship between the blue and G-band brightness can not be explained. In this case the upwelled regions of umbra would still show blackbody flux distribution. Also, the brightness asymmetry observed in sun center side compared to limb side may not be consistent. In conclusion, the location of bright points at the photospheric intersection of the diverging umbral flux tubes seems to be the natural model that accommodates most of the observed features. 

\acknowledgments

One of the authors DPC is grateful The Institute of Space and Astronautical Sciences for the hospitality during this work. We would like to thank Dr. Eugene Parker for stimulating discussions. A part of support comes from NSF under grant ATM 05-48260 and NASA grant NNX 08AQ32G. The data was obtained using Hinode Solar Optical Telescope. Hinode is a Japanese mission developed and launched by ISAS/JAXA, with NAOJ as domestic partner and NASA and STFC (UK) as international partners. It is operated by these agencies in co-operation with ESA and NSC (Norway). 




\clearpage



\begin{figure}
\epsscale{.80}
\plotone{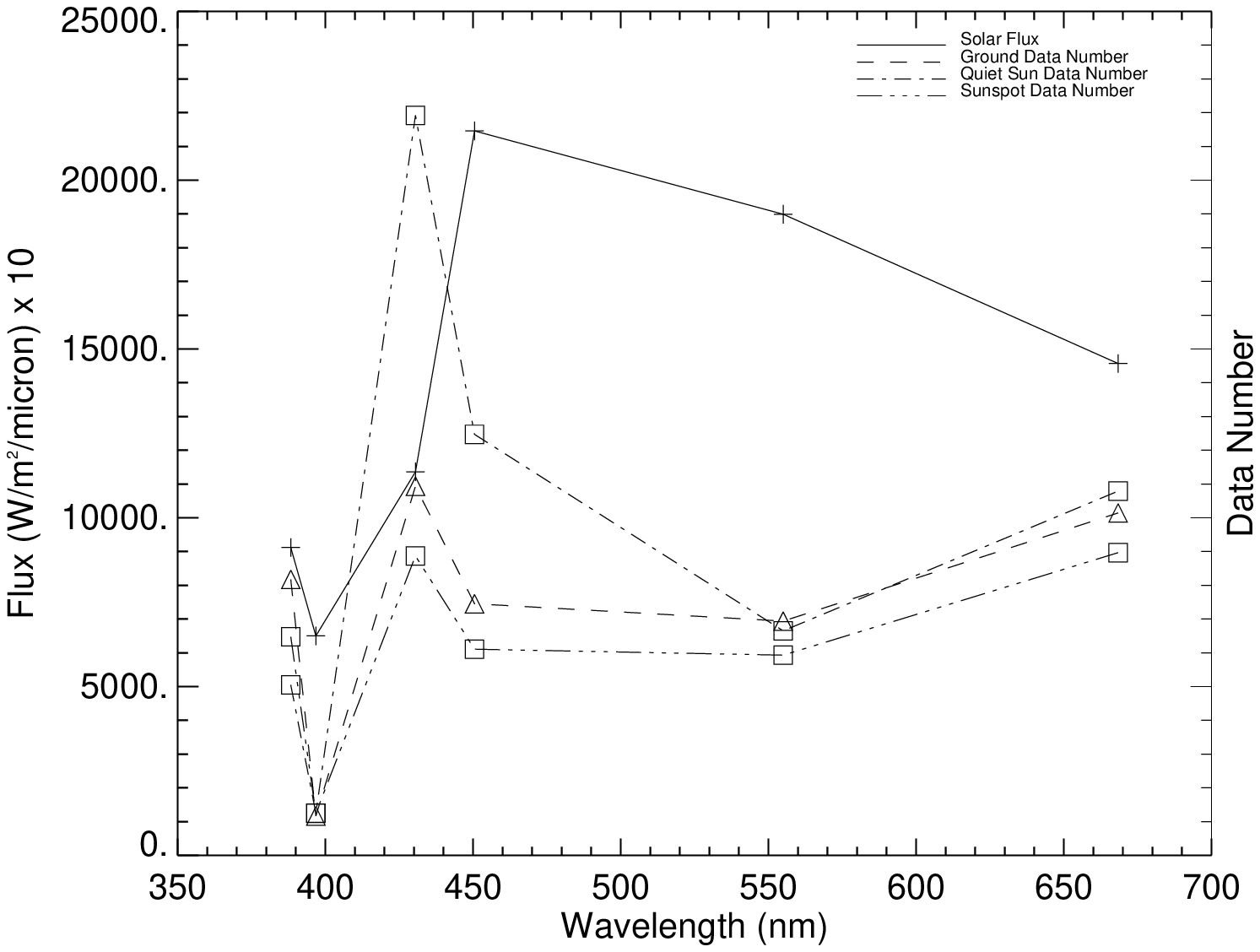}
\caption{Hinode Solar Optical Telescope Broad Band Filter Imager detector Response. The Data Numbers corresponding to the solar flux through different filters, obtained on ground \citep{shi07} and in space, are plotted along with the zero air mass solar spectrum. The dash and dash-dotted curves represent the disk center and quiet sun our side the sunspot respectively. Depending on the disk position of the sunspot the values in dash-dotted corve vary. The left side y-axis refers to the solar flux units and the right side Data Numbers. The Zero Airmass Solar Spectrum can be found at http://rredc.nrel.gov/solar/spectra/am0/ \label{fig1}}
\end{figure}

\clearpage
 
\begin{figure}
\epsscale{1.}
\plotone{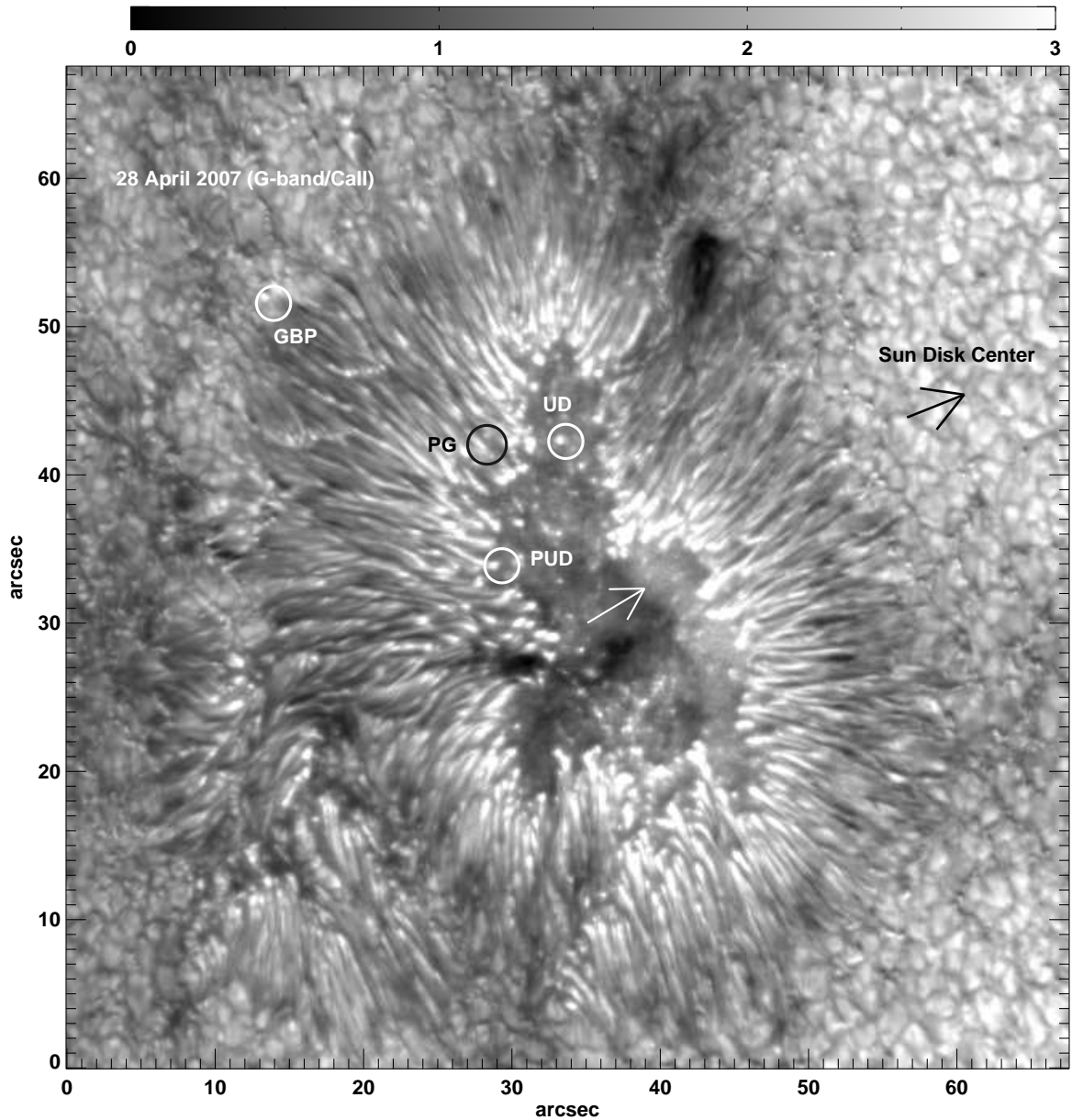}
\caption{The contrast enhanced ratio image of G-band and CaII image of the sunspot. The sun-center is to the right. The umbral dots are seen enhanced. The inner penumbra is brighter than the outer penumbra. The umbral dots near the penumbra are associated with elongated structures. The Umbral Dot (UD), Penumbral Grain (PG), Peripheral Umbral Dot (PUD) and G-Band Bright Point (GBP) are shown in circles. The arrow show the elongated structure associated with umbral dots. The sun-center is to the right.\label{fig2}}
\end{figure}

\clearpage

\begin{figure}
\plottwo{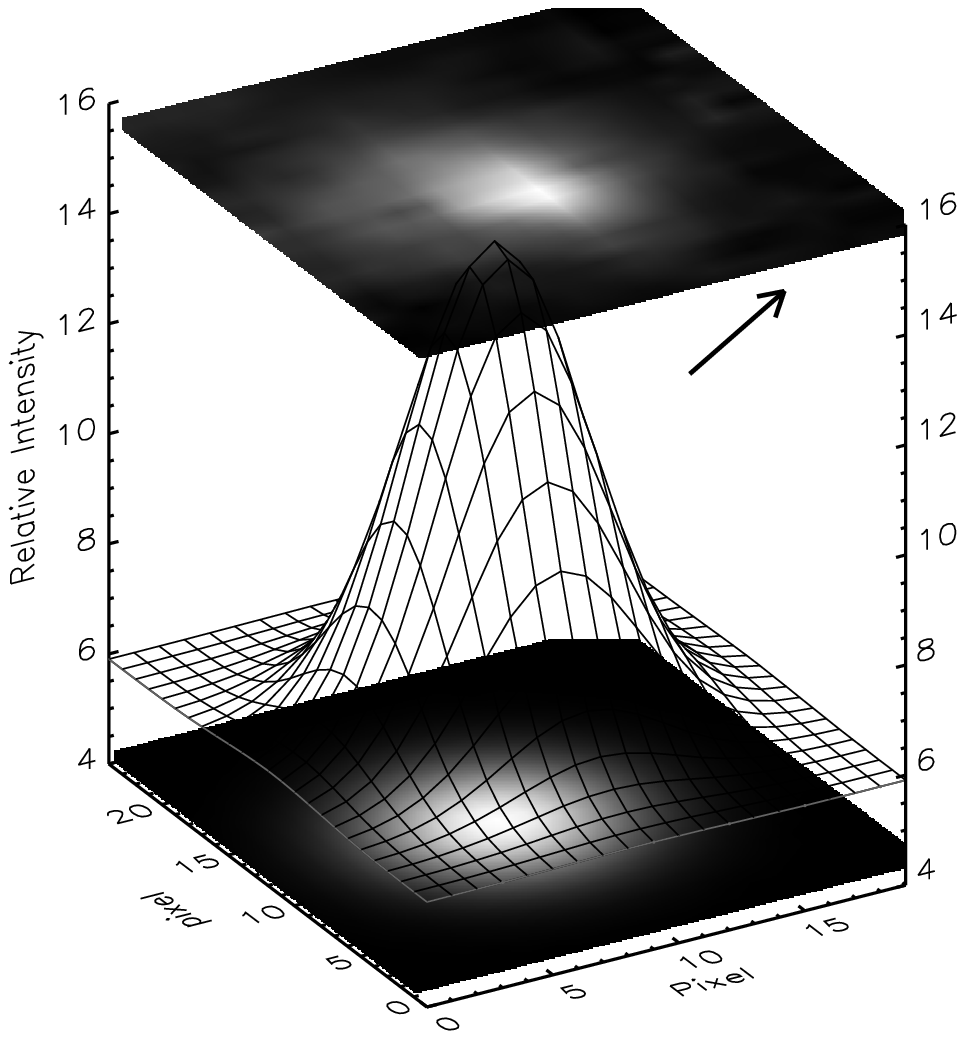}{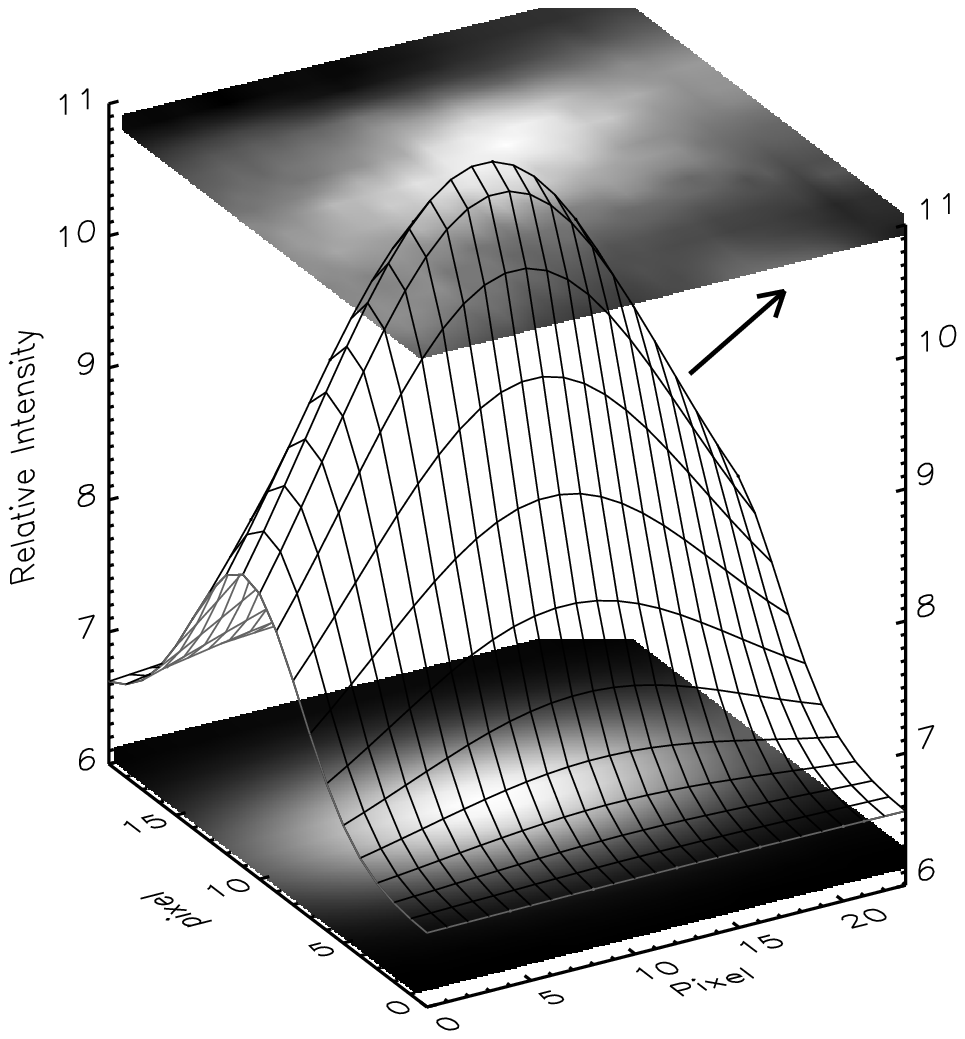}
\caption{(a) Umbral dot, (b) Penumbral grain. The top and bottom images are the sunspot bright point and fitted images. The surface plot represent the two dimensional Gaussian function for the bright points. The arrow mark points towards to the penumbra in (a) and penumbral foot point in (b).\label{fig3}}
\end{figure}

\clearpage

\begin{figure}
\epsscale{.80}
\plotone{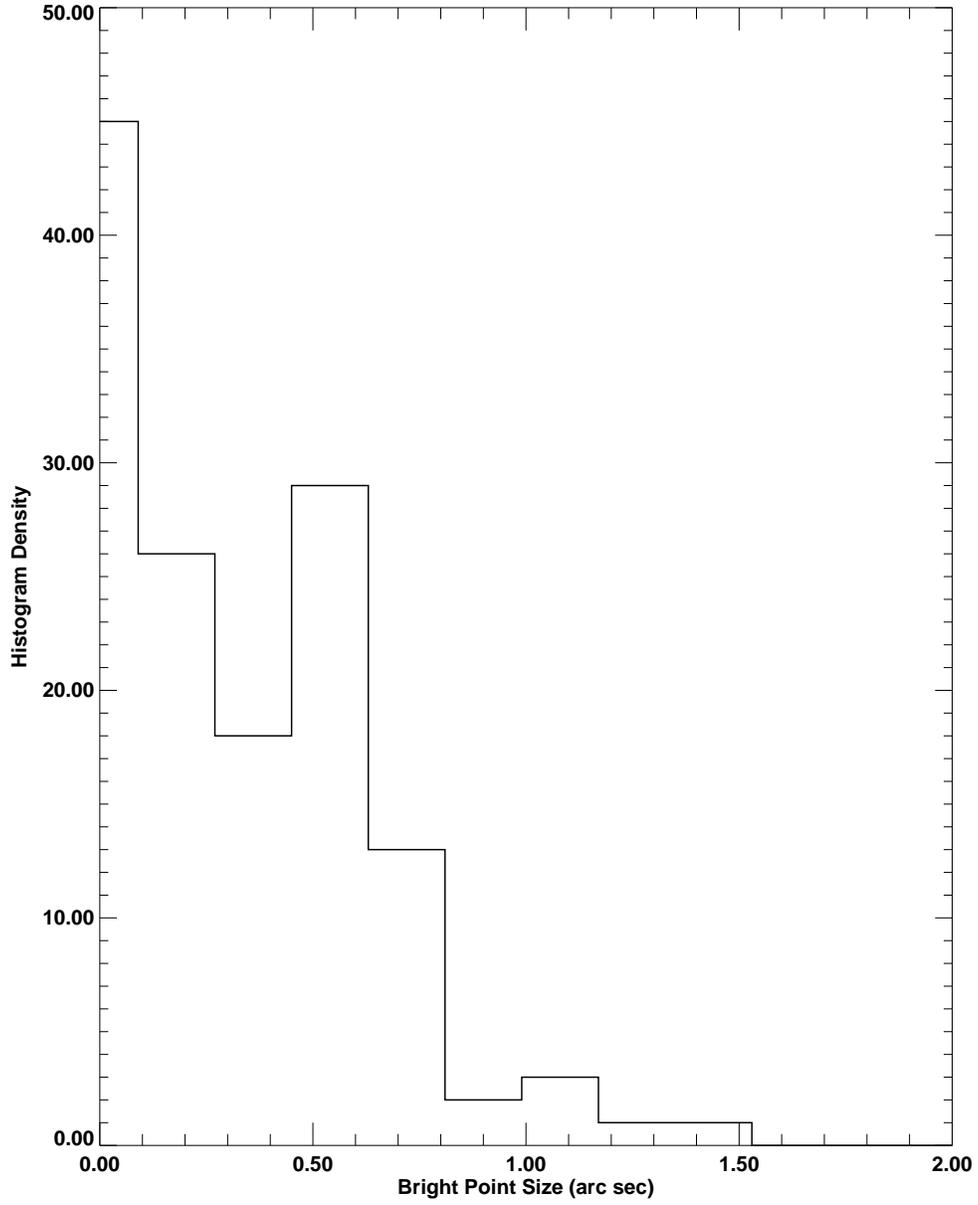}
\caption{The histogram of size (bright point radius) distribution of the sunspot bright points. Umbral dots are smaller than 0.25 arc second, size of peripheral umbral dots (penumbral foot points) peak at 0.5 arc second and penumbral grains are larger. \label{fig4}}
\end{figure}

\clearpage
 
\begin{figure}
\includegraphics[angle=0,scale=.40]{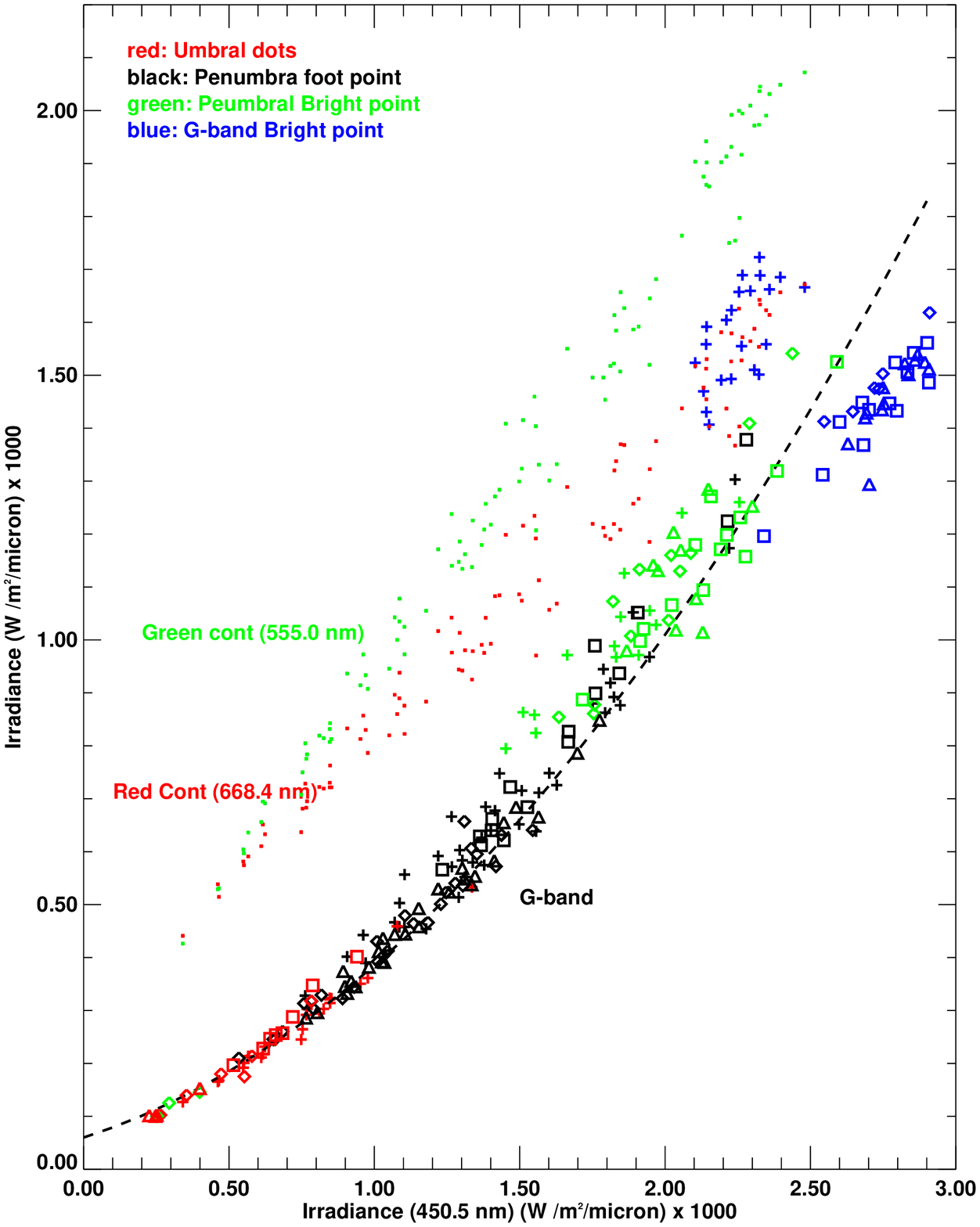}
\includegraphics[angle=0,scale=.40]{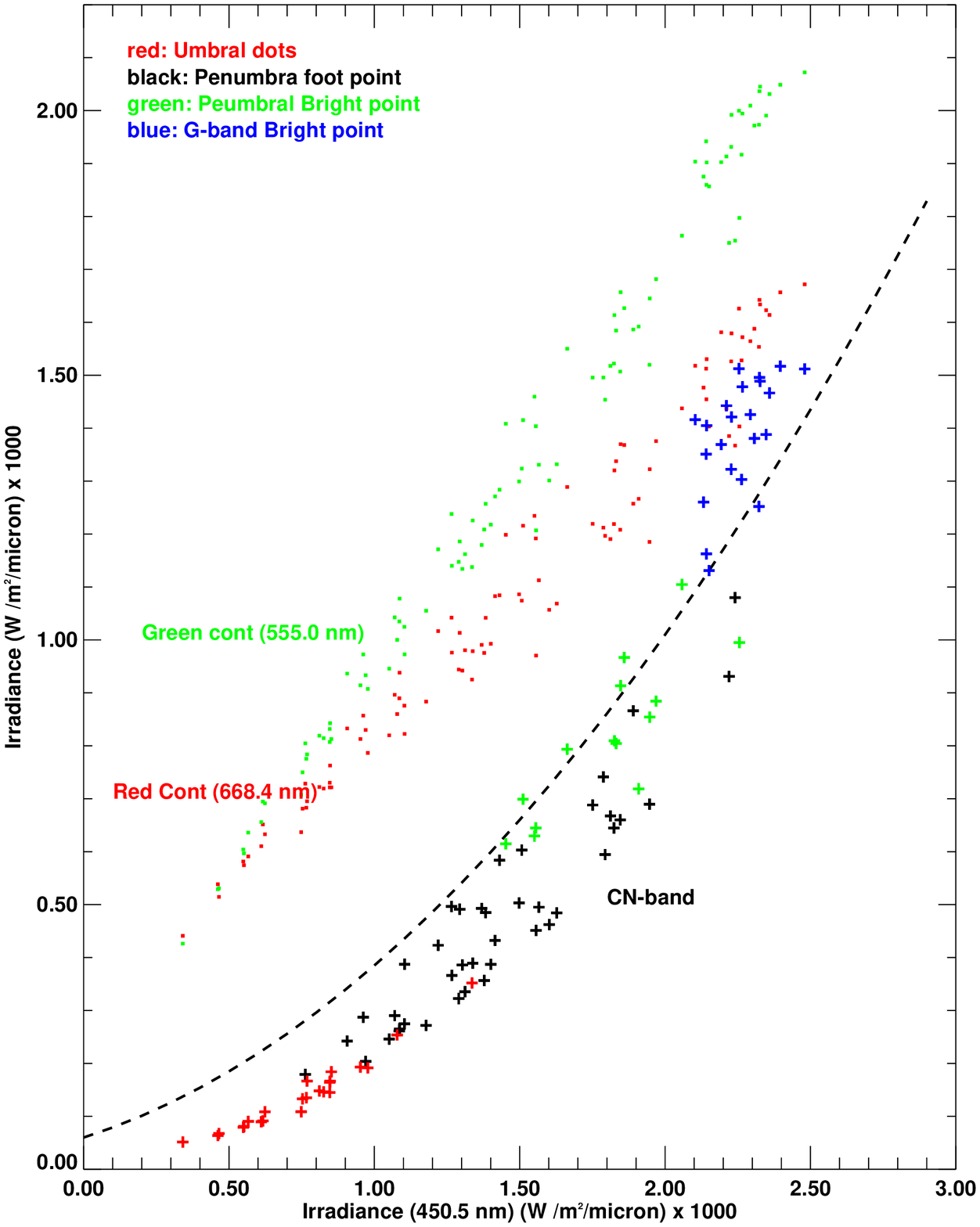}
\includegraphics[angle=0,scale=.40]{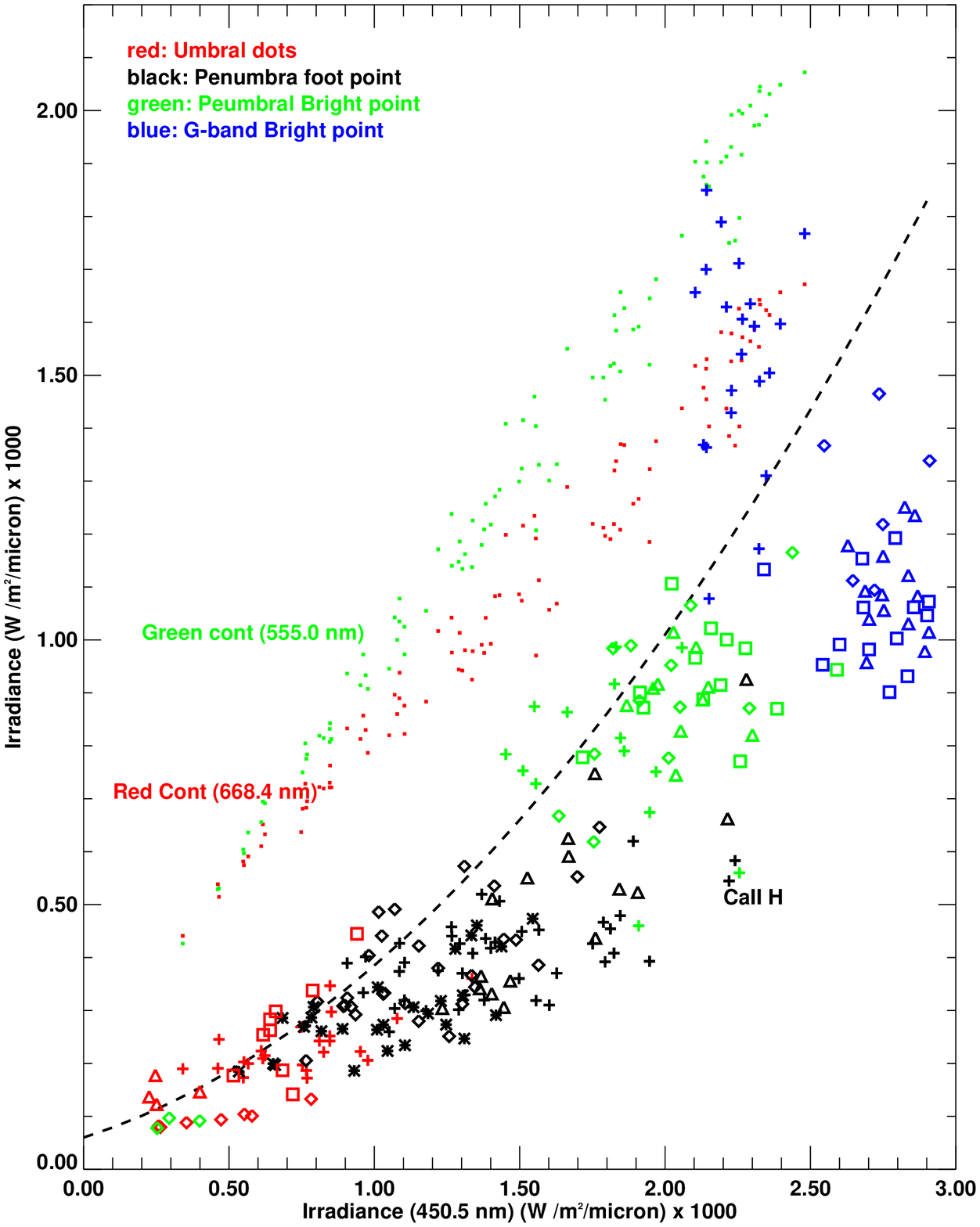}
\caption{The flux of the sunspot bright points through G-band, red and green filters plotted as a function of flux through blue filter. The symbols represent different sunspots. a (top right), b(top left) and c (bottom) figures are the scatter plot for G-band, CN-band and CaII H flux respectively. The small plus symbols represent the continuum flux through SOT green and red filters.}
\end{figure}

\clearpage

\begin{figure}
\epsscale{.80}
\plotone{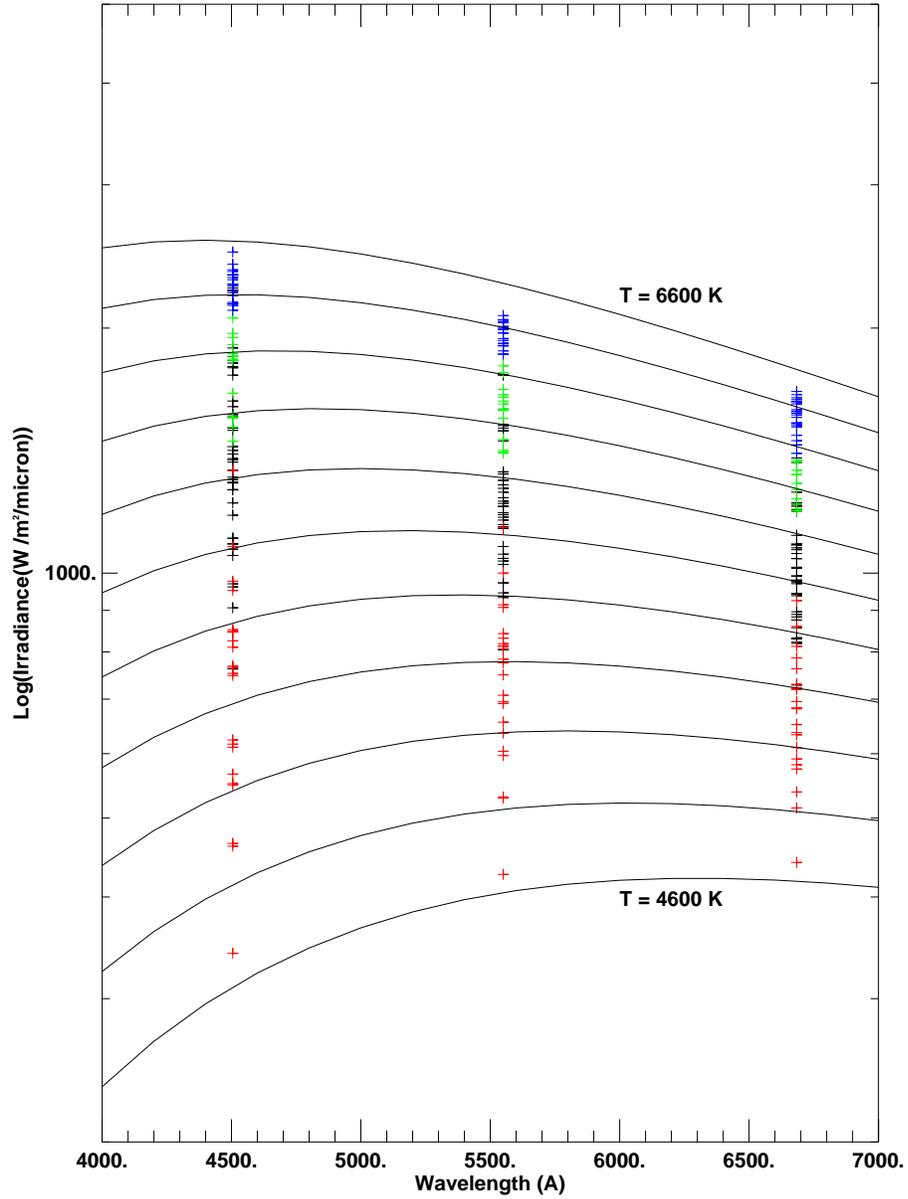}
\caption{Continuum filters superposed with planck functions of temperature range 4600 to 6600 K with intervals of 200 K. The red to blue + sign are umbral dots to circum-sunspot G-band bright points. The temperature vary gradually among the bright points depending on their location in the spot. The umbral dots are cooler than those situated outside the spot.} \label{fig6}
\end{figure}

\clearpage

\begin{figure}
\epsscale{.80}
\plotone{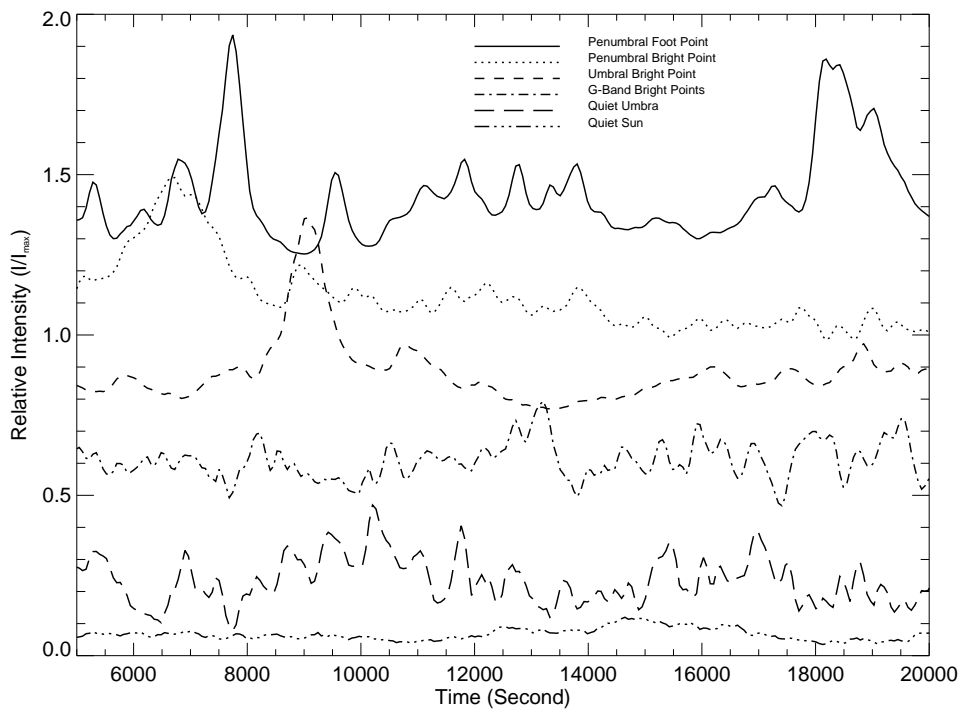}
\caption{The light curves of different bright points in and around the sunspot. The light curves are normalized to the peak intensity (I$_{max}$) of the time series for the individual bright points. The profiles are plotted by shifting 0.5 in y-axis to show them in the same Figure. For each profile the maximum value is 1.0 and the minimum is 0.0. \label{fig7}}
\end{figure}

\clearpage

\begin{figure}
\includegraphics[angle=0,scale=.45]{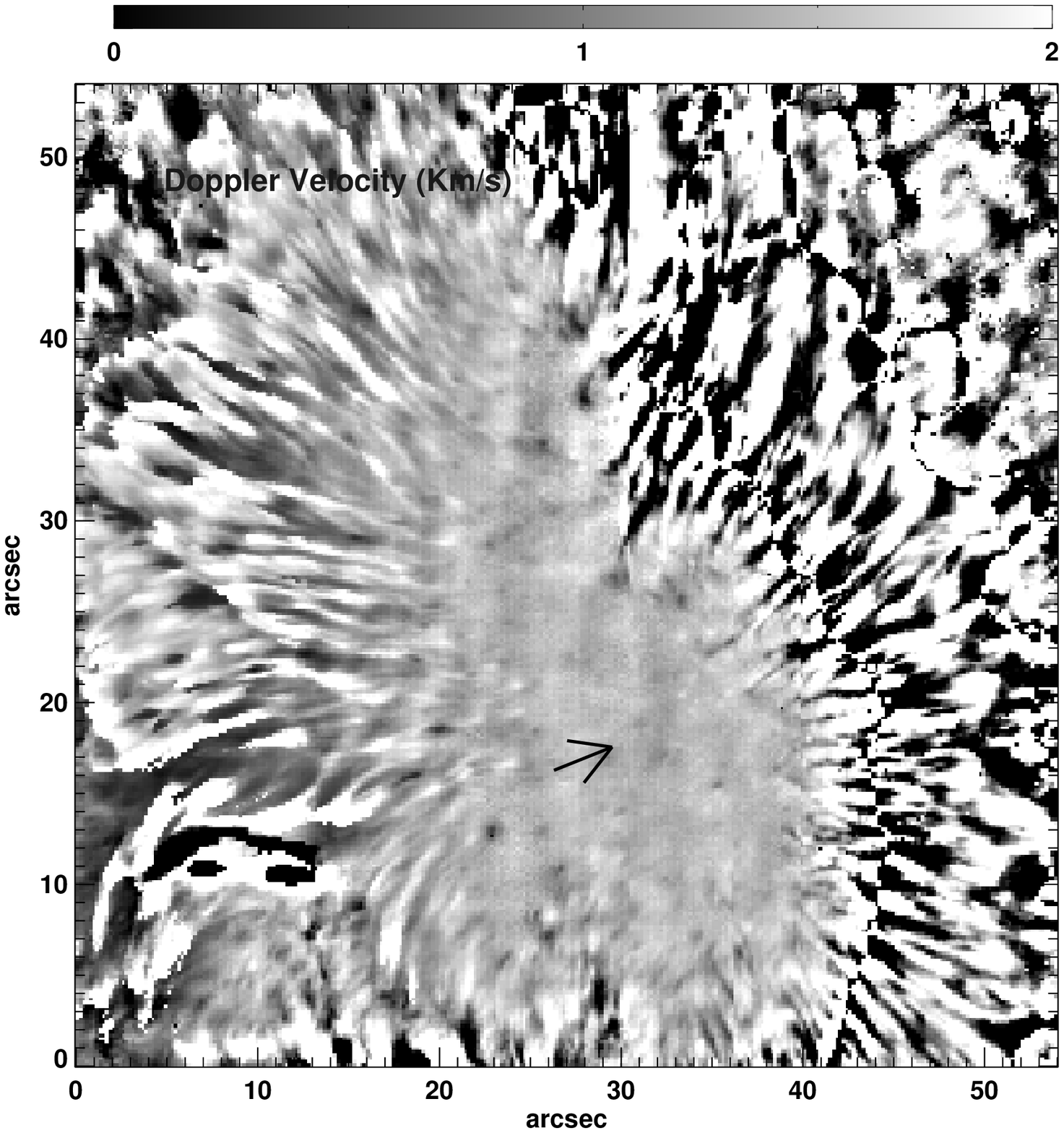}
\includegraphics[angle=0,scale=.45]{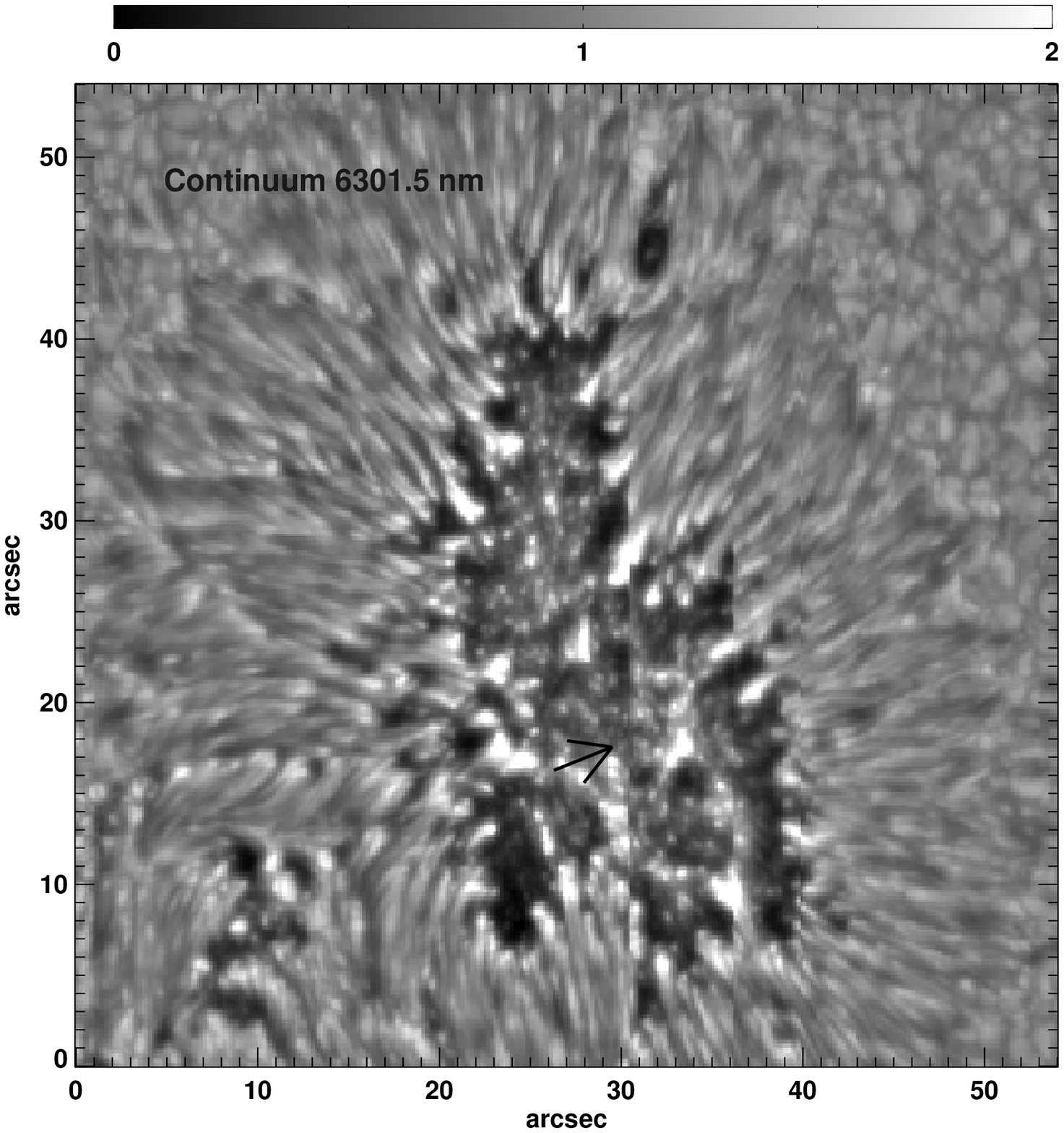}
\includegraphics[angle=0,scale=.45]{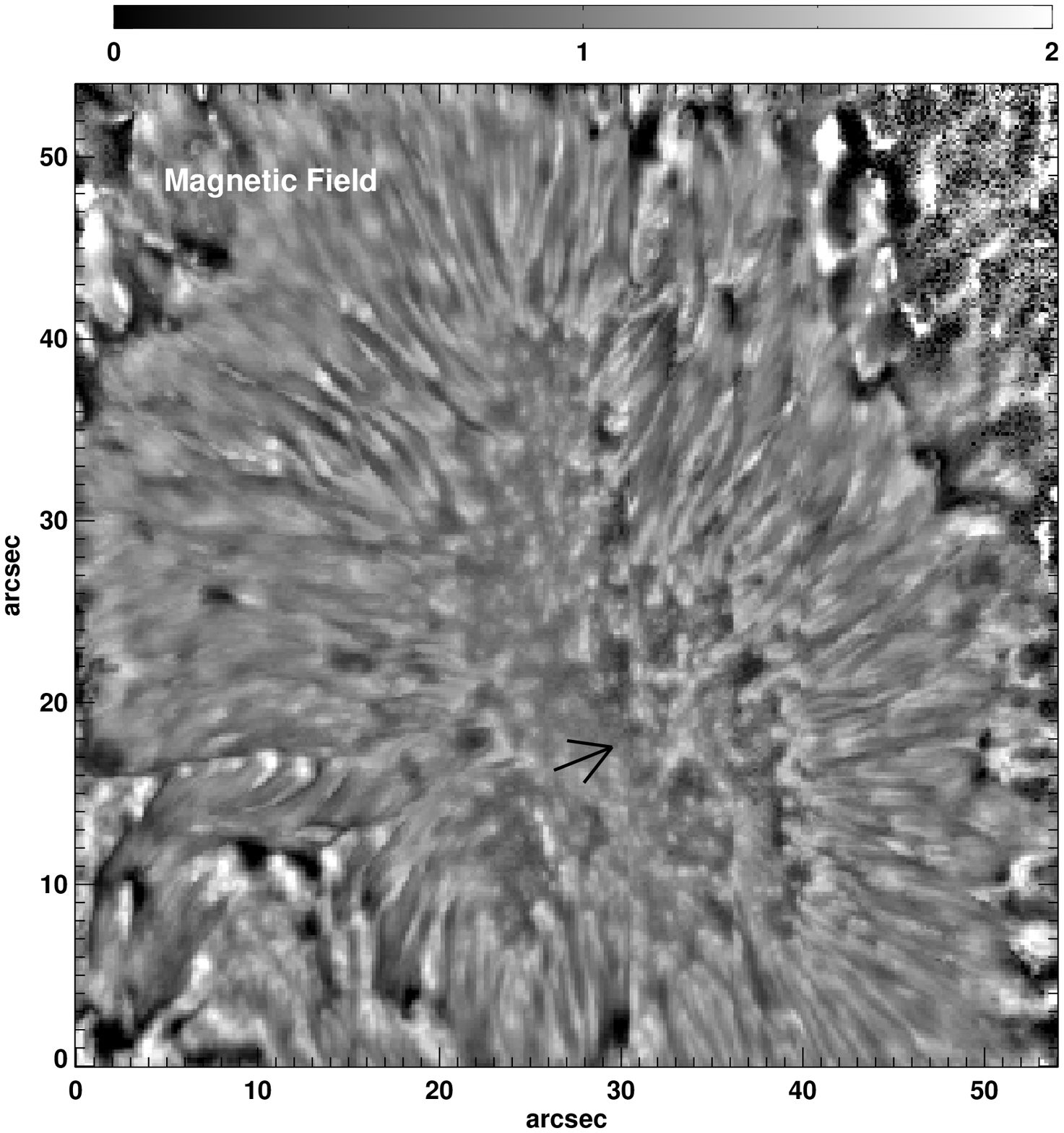}
\includegraphics[angle=0,scale=.45]{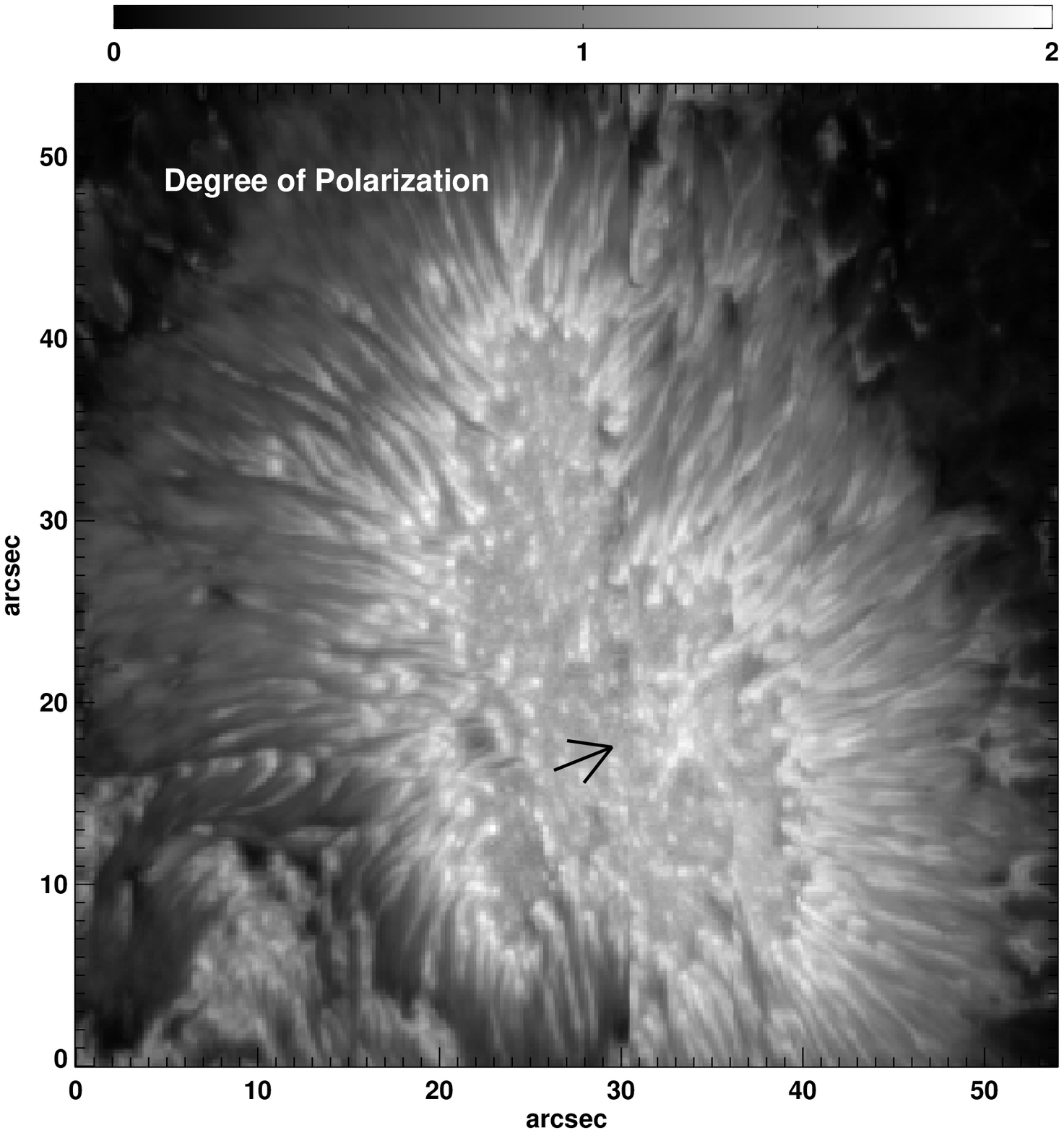}
\caption{(a) Doppler Velocity, (b) Continuum Image, (c) The Magnetic Flux and (d) Degree of Polarization map of the sunspot. The enhancement of magnetic field and degree of polarization corresponding to the bright points are noticed. The Doppler velocity map show narrow channels directed towards penumbra originating from the umbral dots.}
\end{figure}

\clearpage

\begin{figure}
\epsscale{.80}
\plotone{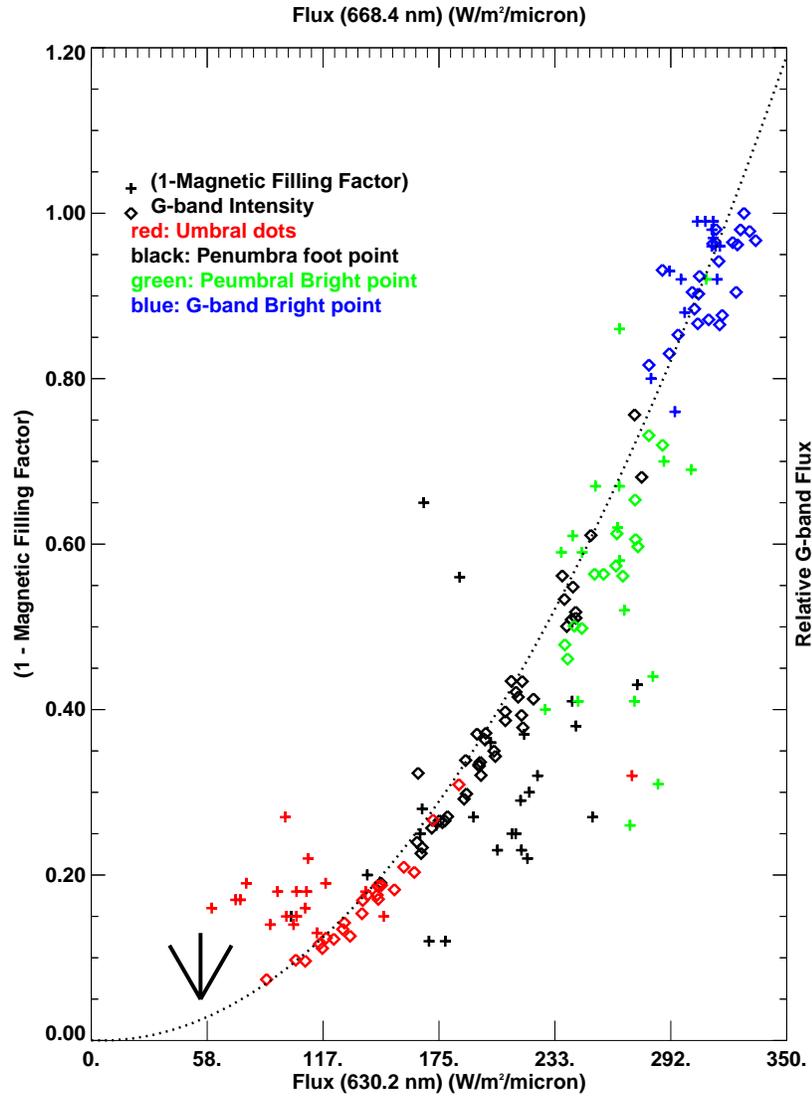}
\caption{Magnetic Filling factor and G-band intensity of sunspot bright points. The left and bottom axis labels is for the magnetic filling factor. The Top and Right axis labels are for the G-band intensity. The bottom arrow indicates that the actual value of the magnetic filling factor of the umbral bright point may be higher.\label{fig9}}
\end{figure}

\clearpage

\begin{figure}
\epsscale{1.0}
\includegraphics[angle=0,scale=.4]{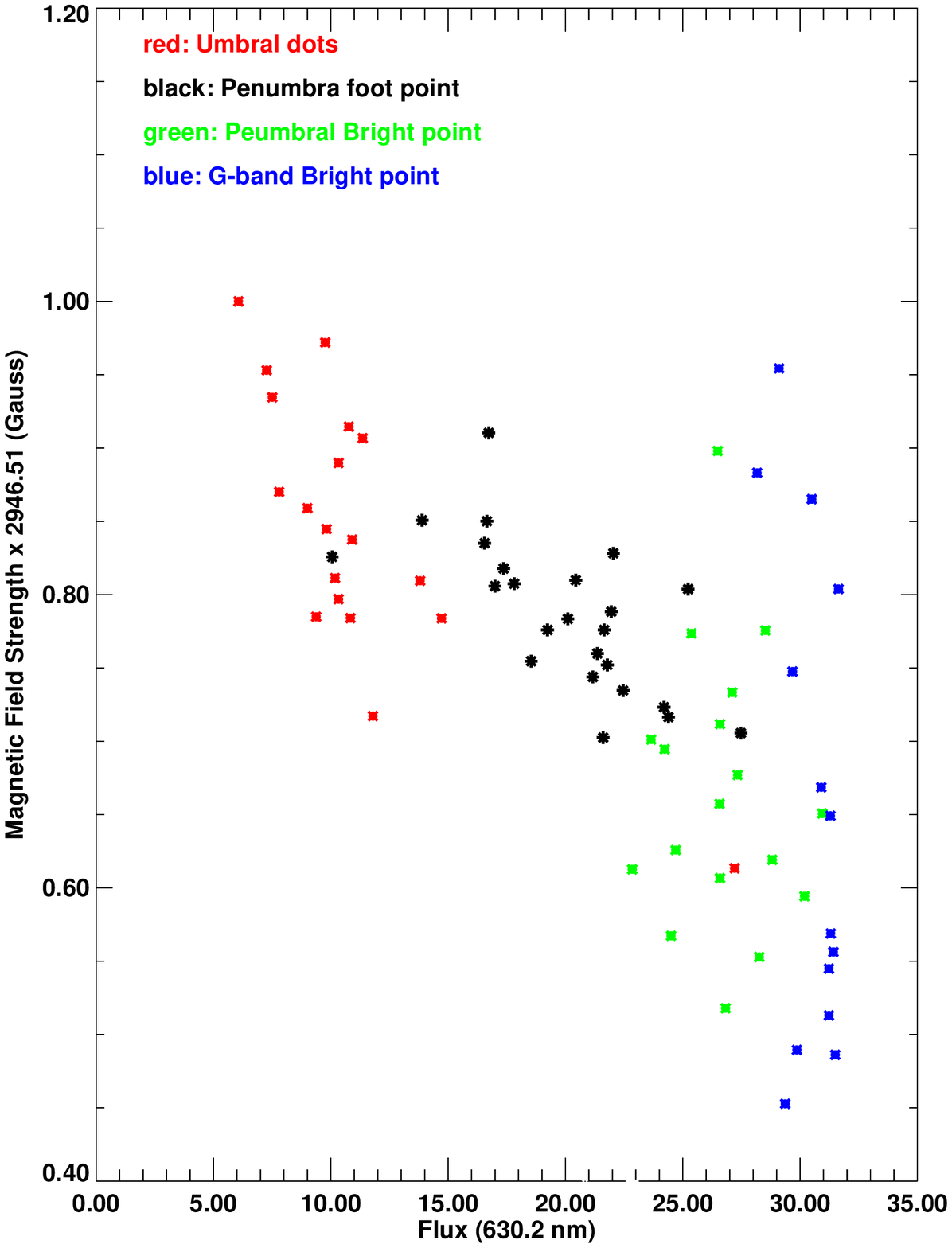}
\includegraphics[angle=0,scale=.4]{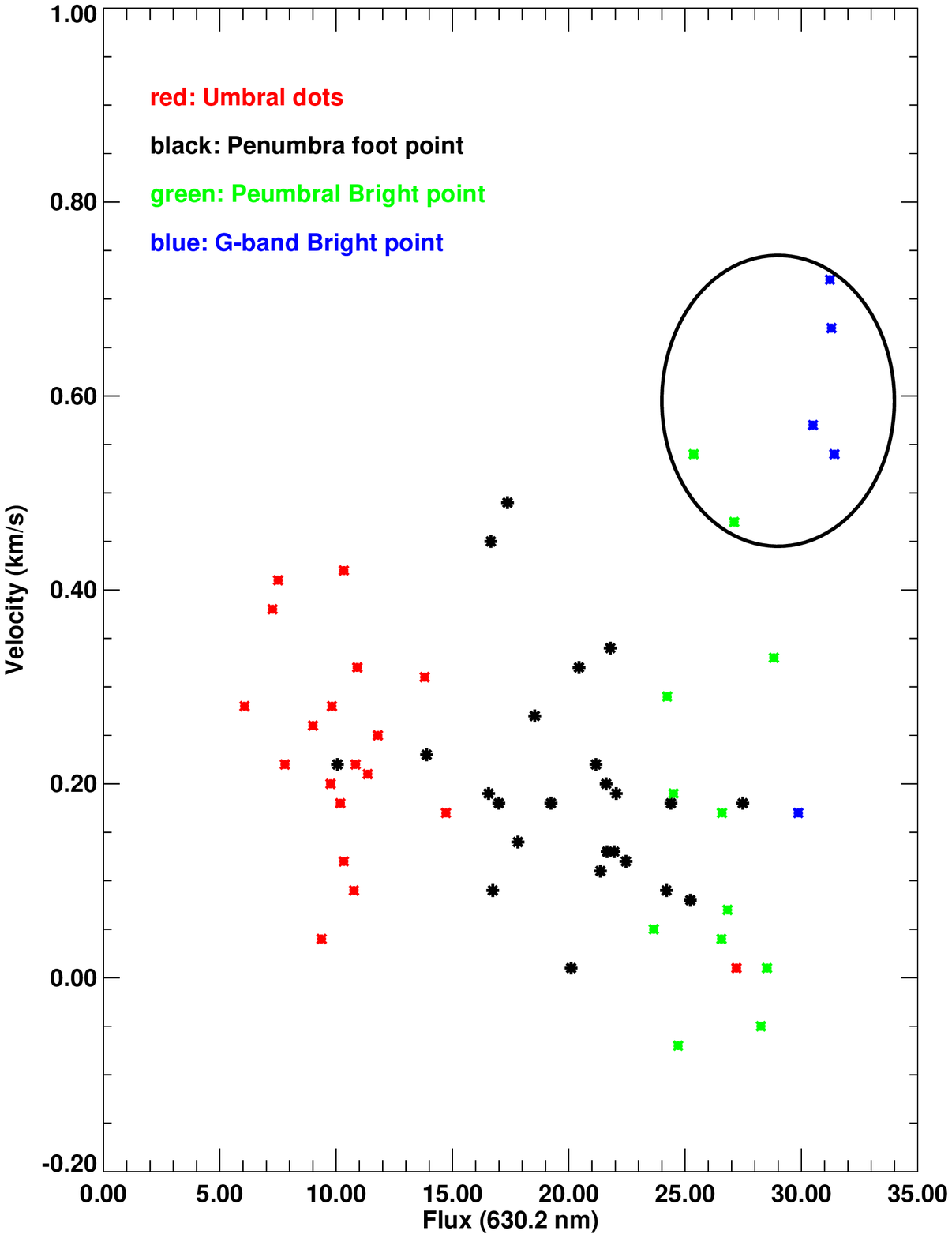}
\includegraphics[angle=0,scale=.4]{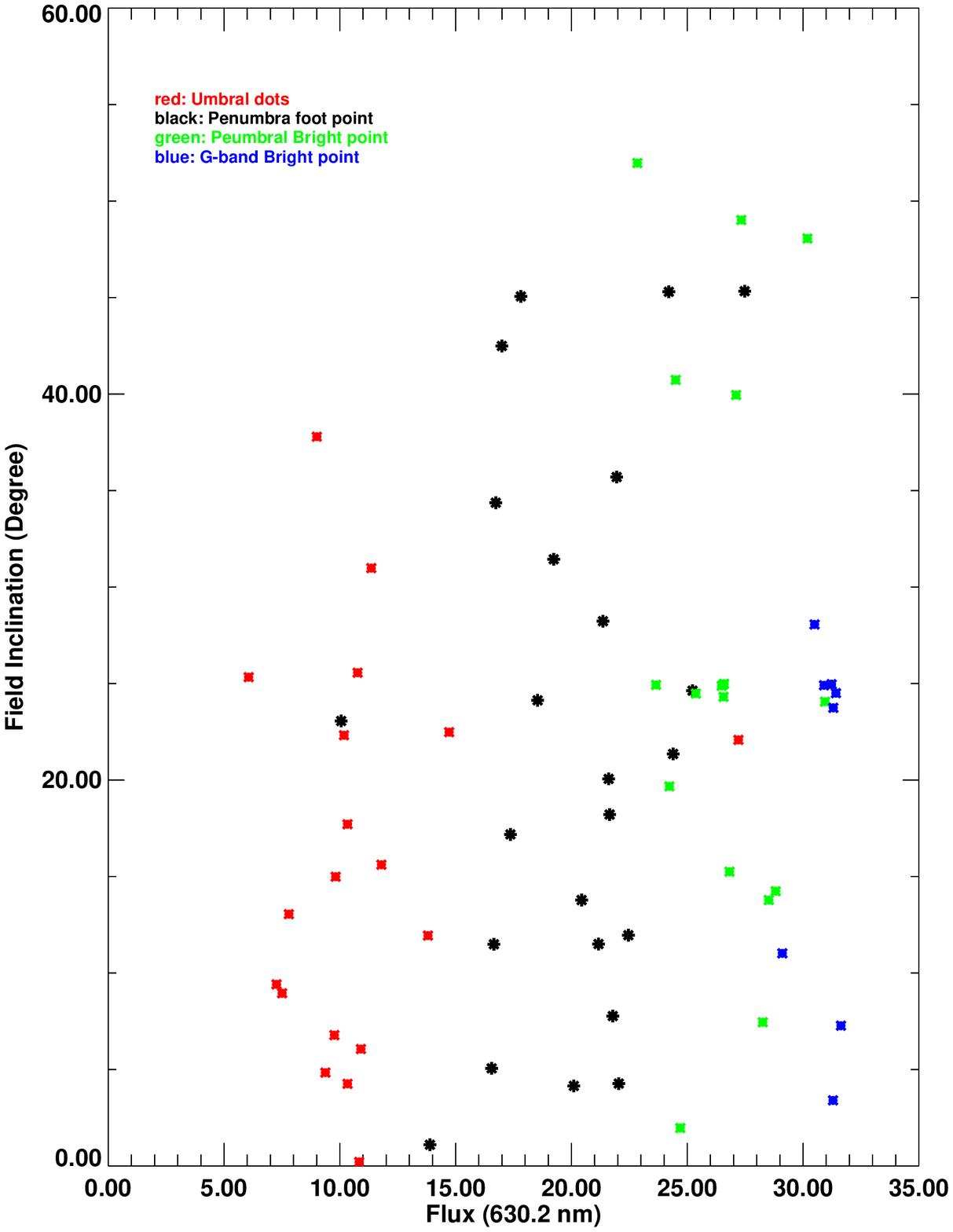}
\includegraphics[angle=0,scale=.4]{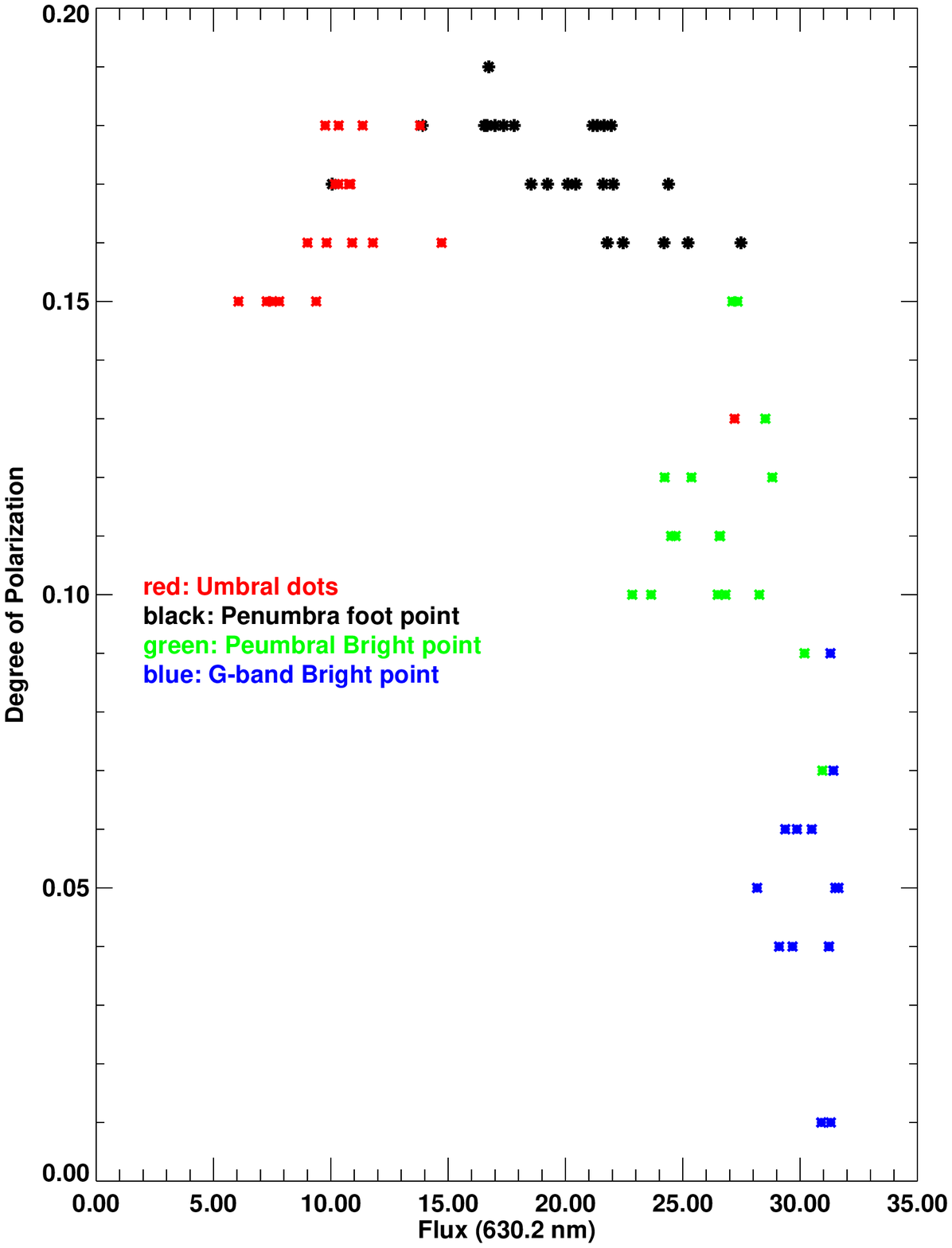}
\caption{(a) Magnetic Field, (b) Doppler Velocity, (c) Magnetic Field inclination angle and (d) Degree of Polarization of the bright points of the sunspot. The enhancement of magnetic field and degree of polarization corresponding to the bright points are noticed. The encircled points in velocity plot showing higher velocity belong to the bright points towards the end of the penumbra and outside the sunspot.}
\end{figure}

\clearpage

\begin{figure}
\epsscale{.80}
\plotone{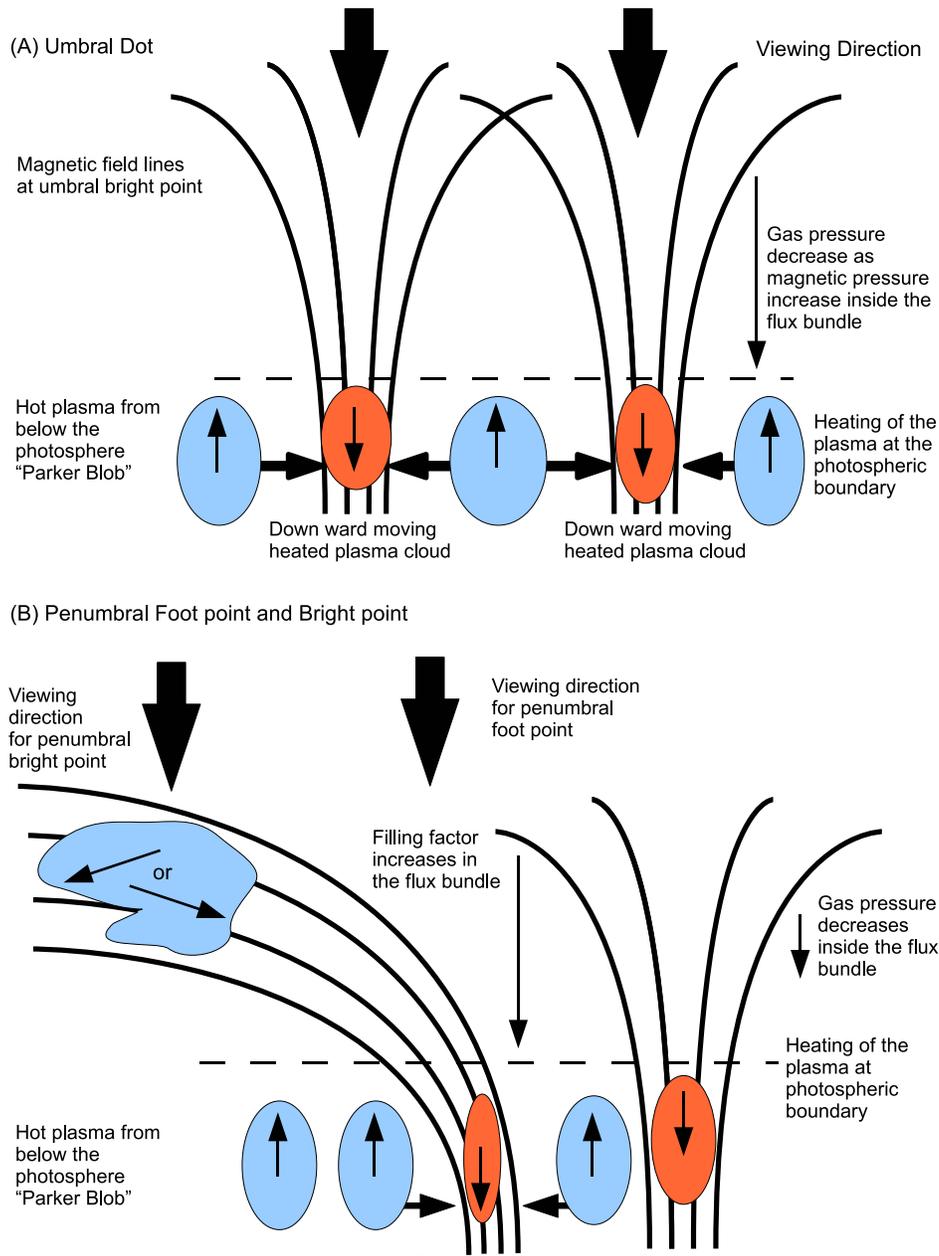}
\caption{The Model. The dark solid lines represent the magnetic field lines. The blue blob are magnetic field free plasma inter-flux tube space. The orange blobs are plasma at the foot point of the flux tubes that are heated by the oscillating field free plasma penetrating from the side or below the sunspots. The arrows on the blobs indicate the direction of their momentary state of motion. The thick arrow indicate the viewing direction of the bright features. The side arrows from the blue blobs indicate the direction of heat flow from non-magnetic material to flux tube material. \label{fig11}}
\end{figure}

\clearpage

\begin{deluxetable}{ccrrrrrrrrcrl}
\tabletypesize{\scriptsize}
\tablecaption{Sunspots used for the study}
\tablewidth{0pt}
\tablehead{
\colhead{POS} & \colhead{Date} & \colhead{Location} & \colhead{Hale Type} & \colhead{Area$^a$} 
 }
\startdata
2006 November 14& 10940 & S10E35 & $\beta \delta$/$\alpha$  & 500        \\
2007 February 28& 10944 & S05W07 & $\beta$ & 640         \\
2007 February 02& 10923 & S06W03 & $\alpha$  & 100     \\
2007 April    28& 10953 & S04W20 & $\beta$ & 290         \\
\enddata
\tablenotetext{a}{in millionth of disk}
\end{deluxetable}


\clearpage








\clearpage
\end{document}